\documentclass[a4paper, 10pt,oneside,number,sort&compress]{elsarticle}
\usepackage{lineno}
\usepackage{graphicx}
\usepackage[latin1]{inputenc}
\usepackage{verbatim}
\usepackage{amsfonts}
\usepackage{color}
\usepackage{epsfig}
\usepackage{setspace}
\usepackage{amstext}
\usepackage{amsbsy}
\usepackage{amscd}
\usepackage{amsgen}
\usepackage{amsopn}
\usepackage{amsxtra}
\usepackage{xspace}
\usepackage{colortbl}
\usepackage{captcont}
\usepackage{subfigure}

\usepackage{footnote}
\makesavenoteenv{tabular}
\usepackage[font=footnotesize,hang]{caption} % per le note

\usepackage{floatrow}% Table float box with bottom caption, box width adjusted to content
\newfloatcommand{capbtabbox}{table}[][\FBwidth]

\usepackage{blindtext}

\usepackage{fancyhdr}
\usepackage{multicol}

\usepackage{pifont}%
\usepackage{geometry}%
\usepackage{fleqn}
\usepackage{txfonts}
\usepackage{hyperref}

\def\Xmax{\ifmmode {X_\mathrm{max}}\else
                   {$X_\mathrm{max}$}\fi\xspace}%

\begin{document}

\begin{frontmatter}

\title{Identification of the primary mass of inclined cosmic ray showers from depth of maximum and number of muons parameters}

\author[INAF]{S. Riggi\corref{cor}}
\ead{simone.riggi@ct.infn.it}\cortext[cor]{Corresponding author.}
\author[USC]{A. Parra}
\author[Roma2]{G. Rodriguez}
\author[USC]{I. Vali\~{n}o} 
\author[USC]{R. V\'{a}zquez} 
\author[USC]{E. Zas}

\address[INAF]{INAF - Osservatorio Astrofisico di Catania, Italy}
\address[USC]{Universidad de Santiago de Compostela, Spain}
\address[Roma2]{Universit\`{a} di Roma II ``Tor Vergata'' and INFN Roma, Italy}

\begin{abstract}
% Text of abstract
In the present work we carry out a study of the high energy cosmic rays mass identification capabilities of a hybrid detector employing
both fluorescence telescopes and particle detectors at ground using
simulated data. It involves the analysis of extensive showers with zenith angles above 60 degrees  
making use of the joint distribution of the depth of maximum and muon size at ground level as mass discriminating parameters. The 
correlation and sensitivity to the primary mass are investigated. Two different techniques - clustering algorithms and neural networks - are adopted to classify 
the mass identity on an event-by-event basis. Typical results for the achieved performance of identification are reported and discussed. The analysis can be 
extended in a very straightforward way to vertical showers or can be complemented with additional discriminating observables coming from different types of detectors.
\end{abstract}


\end{frontmatter}

% main text
\section{Introduction}
\label{IntroductionSection}%
Mass composition analysis is fundamental to understand the features observed in the cosmic ray energy spectrum and to test theoretical models 
concerning the origin and the nature of the primary cosmic ray
radiation at the highest energies. 
Current theories predict different spectra for each mass component so
that 
%Different energy spectra are predicted to be observed at 
%ground by present theories, according to the mass of the primary particle, so 
the knowledge of the energy spectra for every mass component, or at least 
for groups of components, is required in order to discriminate among
the proposed models.\\At low energies ($E<10^{14}$\,eV) cosmic ray composition  
can be measured using direct detection techniques, such as
spectrometers and calorimeters.\\At higher energies, mass measurements
are %generally 
performed with indirect techniques, which make use of shower
parameters sensitive to the primary mass. %Among these, 
The depth at which the longitudinal development has its maximum, \Xmax, and both the number of
electrons $N_{e}$ and muons $N_{\mu}$ 
at ground level are most widely used. An extensive review of 
the adopted detection techniques and mass composition results over the entire energy spectrum range is presented in \cite{UngerKampertReview}. 
\\Two kinds of approaches are generally adopted in composition analysis. Likelihood methods, for instance in \cite{DUrsoICRC09} or unfolding analyses, as in 
\cite{KASCADEAnalysis}, aim to infer the average mass abundances or the energy spectra for different mass components on the basis of a set of parameters 
sensitive to the primary mass, without attempting to establish the
mass of each single event. The %re are however cases in which the 
information of the mass identity on an event-by-event basis would
however be extremely useful and would lead to considerable progress. 
For instance when studying possible correlations of the shower arrival direction with given astrophysical objects it is desirable 
to select events according to their mass. %In other cases 
Some worthwhile studies concern the flux of a given primary,
e.g. protons for the p-Air cross section analysis 
or the search of gamma ray or neutrino events in the hadronic background. In all these situations pattern recognition methods, e.g. neural networks as in 
\cite{Tiba,Ambrosio}, linear
discriminant analysis \cite{NeutrinoAugerPaper}, are typically
employed.
%I do not really understand this below. What is the first kind of
%analysis? 
\\Due to the absence of features strongly correlated to the mass and the presence of stochastic shower-to-shower 
fluctuations, both analyses must necessarily deal with a limited number of mass groups, typically four or five in first kind of analysis and two or three 
in event-by-event studies.
%I do not really understand this above. What is the first kind of analysis? 
\\In this work we study the possibility to use neural networks and clustering algorithms as mass classifier tools on the basis 
of two parameters, the depth of shower maximum and the number of muons, measured with hybrid detectors, such as the Pierre Auger Observatory \cite{PAODesignReport} or
the Telescope Array %Project 
\cite{TADesignReport}. In particular we restrict the analysis to events measured with very high zenith angles ($>$60$^{\circ}$) at energies
above $10^{18}$ eV. For such inclined showers the electromagnetic 
component is strongly absorbed before reaching ground level due to the 
large atmospheric depth traversed and the muon component dominates
the signal measured at the particle detectors. To a rough approximation
only the muon reaches ground level. These events therefore represent a
simple way to measure a parameter reflecting the size of the muon
component in experiments not equipped with dedicated muon detectors. 
This is achieved, as detailed in \cite{GonzaloICRC2011}, by fitting
the measured signals (after subtracting a small contribution due to the
electromagnetic contamination) to a reference parameterization of the 
muon density at ground level obtained from simulations. 
The depth of maximum can be determined from 
the longitudinal profile extracted from the light measurements made 
with the fluorescence 
telescopes (see for instance \cite{UngerProfileReco}).\\The paper is organized as follows. Section \ref{SimulatedDataSection} describes the data set used for this study, 
built from \textsc{Conex} simulations of extensive air showers, and 
addresses the correlation between the two parameters (\Xmax and muon
size) and the sensitivity to the primary mass. In Section 
\ref{ClassificationMethodsSection} a brief description of the designed
neural network and clustering methods is reported. In Section \ref{ResultsSection} 
their application to samples of simulated data and the achieved
classification results are presented. Our main conclusions are 
summarized in Section \ref{SummarySection}.

\begin{figure*}[!th]
\centering%
\subtable[$N_{\mu}$ correction factor]{\includegraphics[scale=0.4]{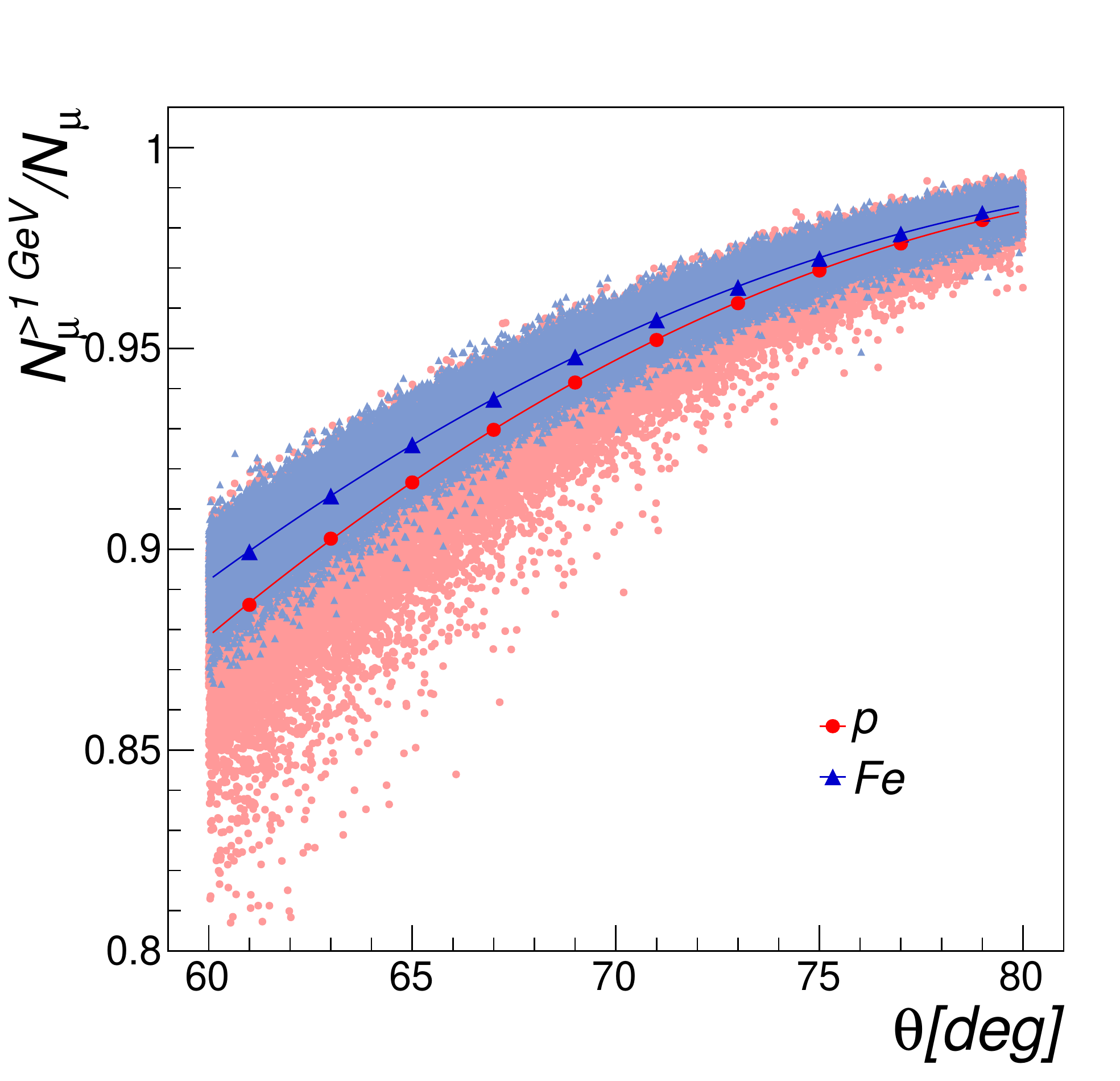}\label{NmuCorrectionFactorFig}}%
\subtable[\Xmax-$N_{19}$]{\includegraphics[scale=0.4]{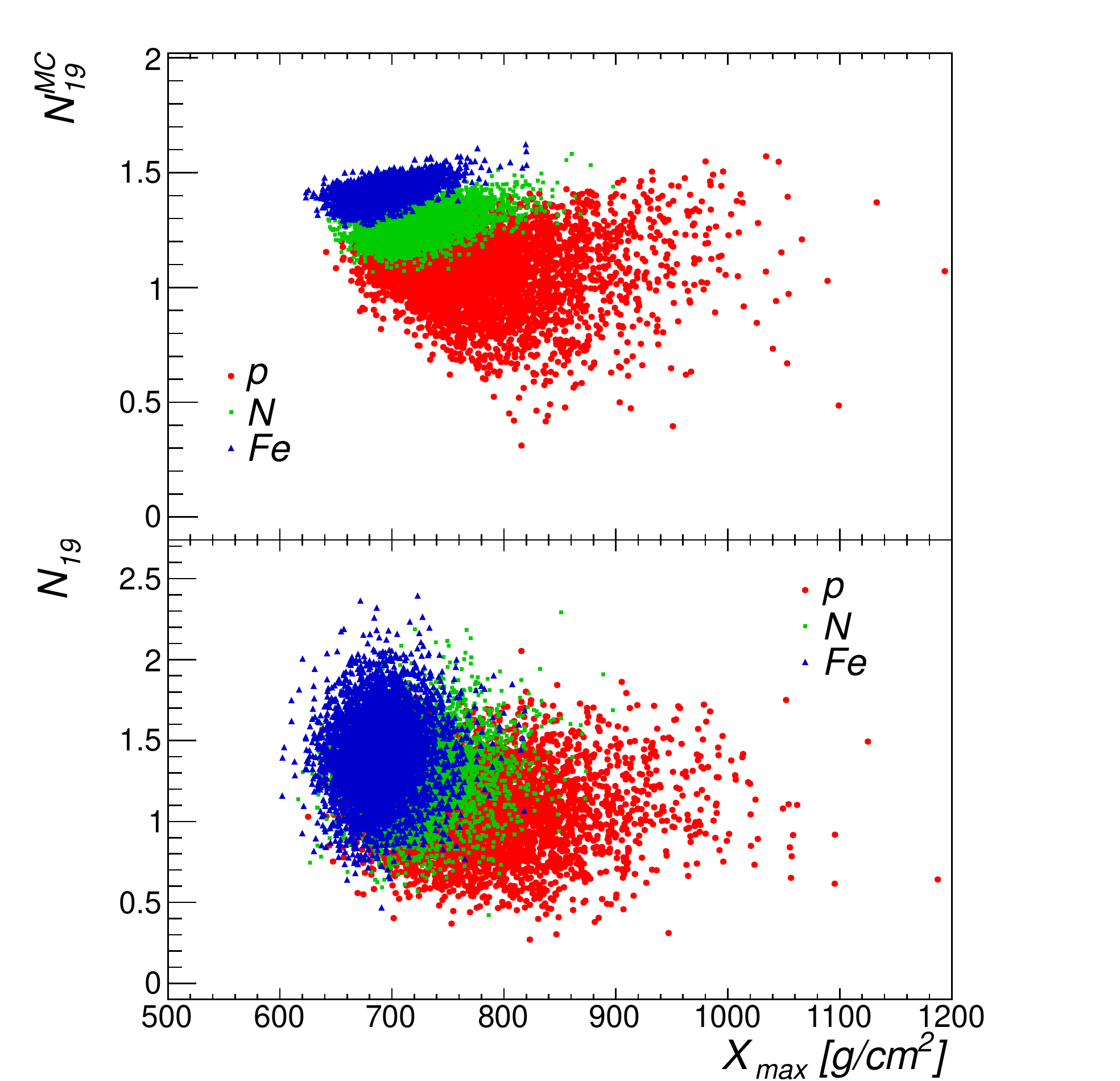}\label{XmaxN19ScatterFig}}%
\vspace{-10pt}%
\caption{Fig. \ref{NmuCorrectionFactorFig}: $N_{\mu}$ correction factor for proton (in red) and iron (in blue) \textsc{Aires} simulations.
Fig. \ref{XmaxN19ScatterFig}: Scatter plot of the \Xmax-$N_{19}$ parameters for proton (red), nitrogen (green) and iron (blue). The bottom panel
refers to the case of detector resolution applied.}%
\label{DataSampleFig}
\end{figure*}

\section{Simulated data}\label{SimulatedDataSection}

\subsection{Simulation strategy}
The present study is based on a sample of simulated showers generated with the \textsc{Conex} tool \cite{Conex1,Conex2} using \textsc{Qgsjet01} \cite{QGSJET01Paper} 
as high energy hadronic interaction model. \textsc{Conex} is an hybrid code, based on the combination of a Monte Carlo strategy and the analytic solution of the 
cascade equations, thus allowing only a one-dimensional simulation of the shower. It requires considerably shorter CPU times with respect to the ones needed by typical 
three-dimensional codes, such as \textsc{Corsika} \cite{CORSIKAPaper} or \textsc{Aires} \cite{AIRESPaper}. This feature makes \textsc{Conex} particularly suitable for applications 
involving data observed with fluorescence detector telescopes.\\Proton, nitrogen and iron primaries have been simulated according to a flat energy spectrum 
from $10^{18}$ eV to $10^{20}$ eV and a $dN/(d\cos\theta)\propto\cos\theta$ zenith angle spectrum from 60$^{\circ}$ to 80$^{\circ}$. About 10$^{6}$ events are 
available per each primary nucleus.\\In order to perform a composition study, a set of parameters sensitive to the primary mass is needed. In this work we assumed
the depth of shower maximum \Xmax and an estimator of the total number of muons $N_{\mu}$ reaching ground level for inclined showers, denoted as $N_{19}$. 
While \Xmax is directly available in \textsc{Conex} simulations, the
latter has to be extracted from the simulated muon profile assuming a
given observation depth. For our analysis we fixed such depth to 1450 m, roughly corresponding to the altitude of the Pierre Auger Observatory.\\Unfortunately \textsc{Conex} 
provides the muon profile only for muons above 1 GeV, therefore we need to employ full \textsc{Aires} simulations to derive an empirical correction factor to
be applied to $N_{\mu}$. In Fig. \ref{NmuCorrectionFactorFig} we display the ratio between the number of muons above 1 GeV $N_{\mu}^{>1 GeV}$ and the total
number of muons $N_{\mu}$ as a function of the shower zenith angle. Proton results are shown in red dots while iron in blue triangles.
As one can see the effect of the muon energy cut is not particularly
relevant for inclined showers. The fraction of unaccounted muons is
below 10\% muons when using \textsc{Conex}. 
An empirical parameterization as a function of the zenith angle is
derived and shown in Fig. \ref{NmuCorrectionFactorFig} with solid colored lines 
\footnote{The energy dependence of the correction factor was also investigated and a negligible dependence was found.}. 
After this correction the simulated $N_{\mu}$ is divided by the number of muons 
of the reference parameterization of the muon density at ground,
obtained for this location at a conventional energy of $10^{19}$ eV 
\cite{GonzaloICRC2011}, to obtain the relative muon size with respect
to the reference energy, $N_{19}^{MC}$. As the number of muons in the reference
parameterization depends on the zenith angle, an estimator 
$N_{19}^{MC}$ is obtained which is basically independent of the zenith
angle.\\To take into account the reconstruction effects, the simulated 
$\Xmax^{MC}$ and $N_{19}^{MC}$ have been smeared with the 
detector resolution. In this work we assume the resolution reported by 
the Pierre Auger Observatory in \cite{GonzaloICRC2011,XmaxPRL}. Typical values for 
$\sigma_{\Xmax}$ are $\sim$27 g/cm$^{2}$ at 10$^{18}$ eV improving at
$<$20 g/cm$^{2}$ above 10$^{19}$ eV. 
The detector resolution $\sigma_{N_{19}}$ has been found 
of the order of 25\% at 10$^{18.5}$ eV improving at $\sim$12\% above 10$^{19}$ eV. 
In Fig. \ref{XmaxN19ScatterFig} we present a scatter plot of \Xmax and 
$N_{19}$ at a fixed energy of 10$^{19}$ eV for the three available
primaries with the upper (lower) panel ignoring (accounting for) detector
resolution as modelled.

\subsection{Correlation of the parameters and sensitivity to primary mass}
\begin{figure*}[!th]
\centering%
\subtable[$\Xmax-N_{19}$ correlation]{\includegraphics[scale=0.4]{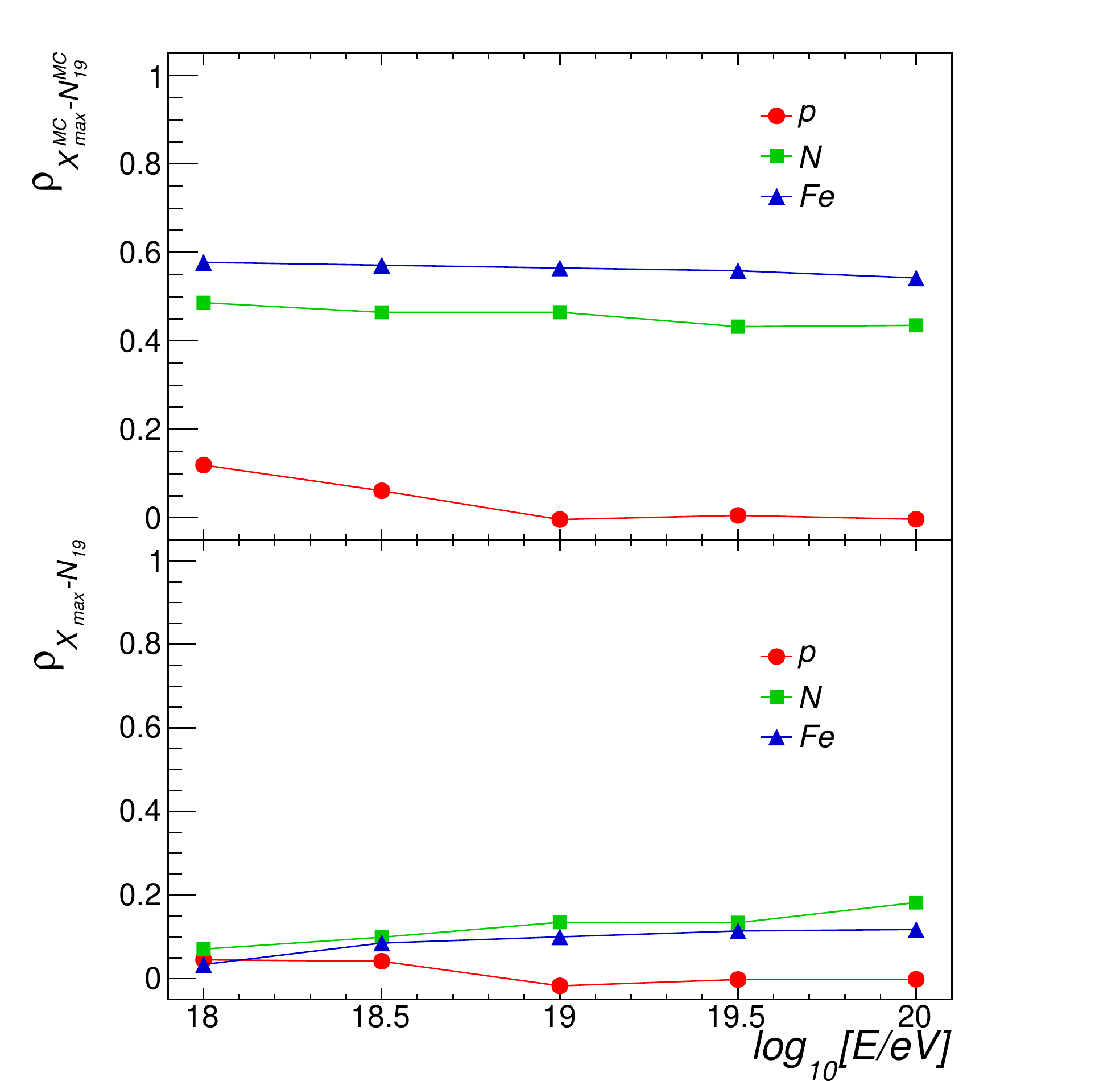}\label{XmaxN19CorrelationFig}}%
\subtable[$\Xmax-N_{19}$ Fisher factor]{\includegraphics[scale=0.4]{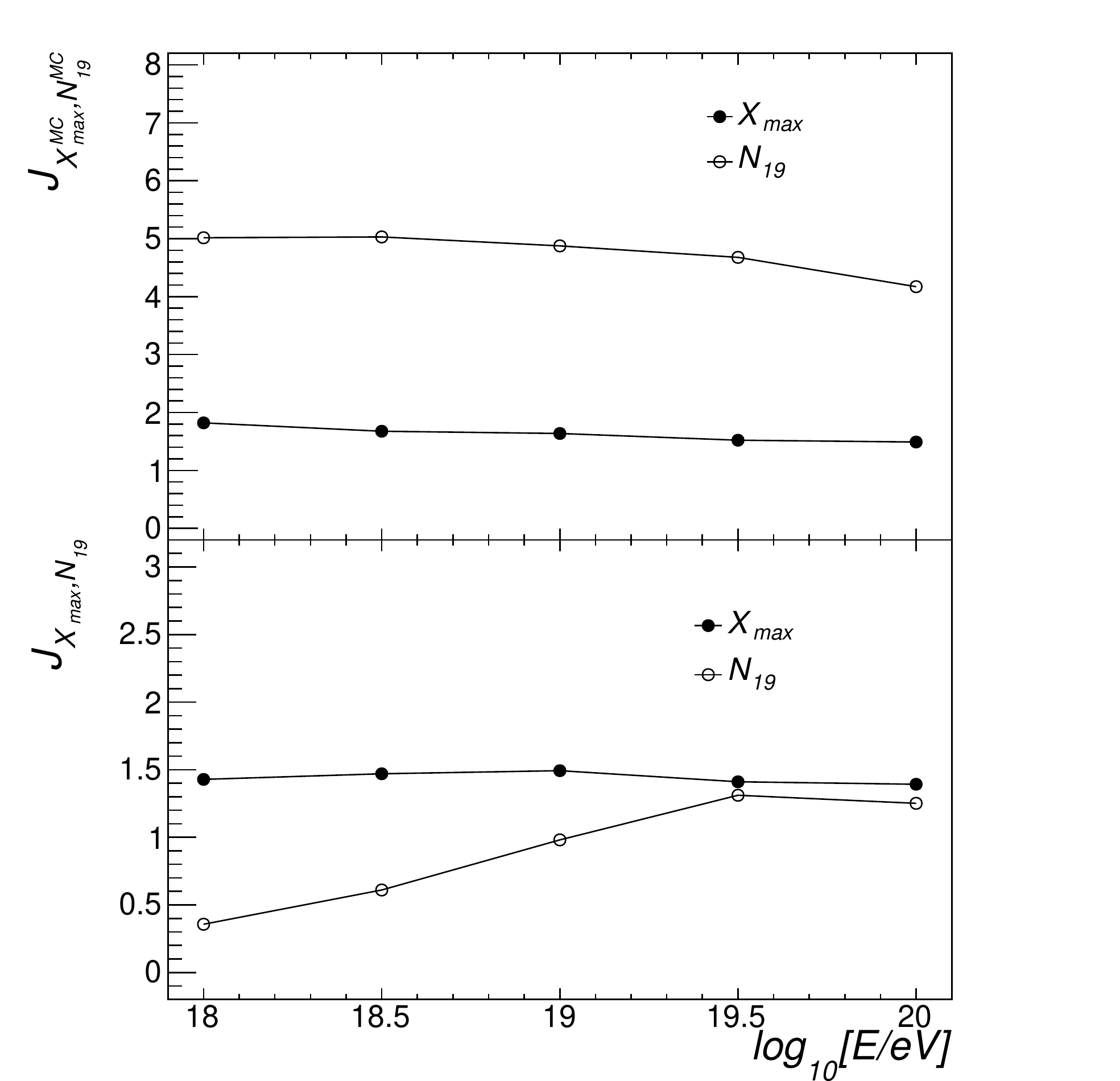}\label{XmaxN19FisherFactorFig}}%
\vspace{-10pt}%
\caption{Fig. \ref{XmaxN19CorrelationFig}: Correlation coefficient of 
$\Xmax$-$N_{19}$ as a function of the primary energy for proton (red), nitrogen (green) and iron (blue). 
Fig. \ref{XmaxN19FisherFactorFig}: Fisher coefficient of the $\Xmax$
and $N_{19}$ parameters with respect to proton and iron primaries. 
Bottom panels take into consideration detector resolution effects.}%
\label{DataSampleFig}
\end{figure*}

It is worth investigating the correlation of the two parameters and evaluating an estimator of the mass sensitivity for both. 
This has been studied in detail for vertical showers in \cite{YounkRisse}. Here we report the results obtained for inclined showers.\\
In Fig. \ref{XmaxN19CorrelationFig} we report the Pearson correlation coefficient of both parameters for the three available nuclei, 
the bottom panel accounting for detector resolution. As 
can be seen the two parameters are almost independent for proton
primaries while for heavier 
nuclei the correlation amounts to 0.5 for nitrogen and 0.6 for iron. 
In the bottom panel the effect of the detector resolution is visible
and the correlation 
observed for nuclei is sensibly reduced.\\To estimate which of the two 
parameters offers the best separation between the proton and iron
distribution we computed the Fisher coefficient $J$ for $\Xmax$ and $N_{19}$:
\begin{equation}
J_{j}= \frac{ (m_{j}^{Fe}-m_{j}^{p})^{2}}{[\sigma_{j}^{Fe}]^{2}+[\sigma_{j}^{p}]^{2}}\;\;\;\;\; j=\Xmax, N_{19}
\end{equation}
where $m_{j}$ and $\sigma_{j}^{2}$ are the sample mean and variance of
the proton/iron distributions of the parameter $j$. The computed quantity 
provides a measure of power to separate two populations of proton and
iron: large values of $J$ correspond to a good discrimination
power. In Fig. \ref{XmaxN19FisherFactorFig} we report the Fisher coefficient 
as a function of energy. As can be seen the number of muons estimator
has a better discrimination power than $\Xmax$. However, the reconstruction of the
$\Xmax$ parameter, especially at low energies, is much better than that of $N_{19}$. This
results in the increased sensitivity of $\Xmax$ when applying the resolution to the data in the bottom panel. At the highest energies both
parameters are almost equally powerful.

\section{Analysis method}\label{ClassificationMethodsSection}
Two different classification methods are used in this work, one based on the neural network technique and the other on clustering algorithms.
In the following sections we briefly describe the two approaches and the strategies adopted to perform the analysis. 

\subsection{The Clustering approach}\label{ClusteringSection}
A simple clustering algorithm, the well-known \textit{k-Means} \cite{KMeansPaper}, is employed for the classification task. 
It starts assuming an initial set of $K$ cluster centroids, which are
then iteratively moved to determine the partition of the data that minimizes a specified square 
error function $E^{2}$. We assumed the Euclidean distance as measure and the following squared error function:
\begin{equation}
E^{2}= \sum_{k=1}^{K}\sum_{j=1}^{n_{k}}\sum_{i=1}^{N}(\mathbf{x}_{ij}-c_{ik})^{2}
\end{equation}
where $x_{ij}$ is the $i_{th}$ data vector (j=$\Xmax$,$N_{19}$) and
$c_{ik}$ are the coordinates of the $K$ cluster centroids.\\The data sample were first 
preprocessed to have zero mean and unit variance and then divided into
three smaller subsets. One of these subsets is used for the cluster training phase in which
the $K$ clusters determined are labelled on the basis of the Monte
Carlo information of the primary mass. The validation subsets are
mainly used to choose the optimal number of clusters,
while the last subset if finally used to test the partition and report
the performance of the method.
As such algorithm is known to depend on the initial choice of the
centroids, the algorithm was run over the three subsets assuming each 
time 100 different initial partitions and retaining the best one in
terms of the total error. A number of clusters 
$K\ge$ 20 has been found to provide optimal
results for our classification problem.
\begin{figure*}[!th]
\centering
\subtable[Clustering output]{\includegraphics[scale=0.3]{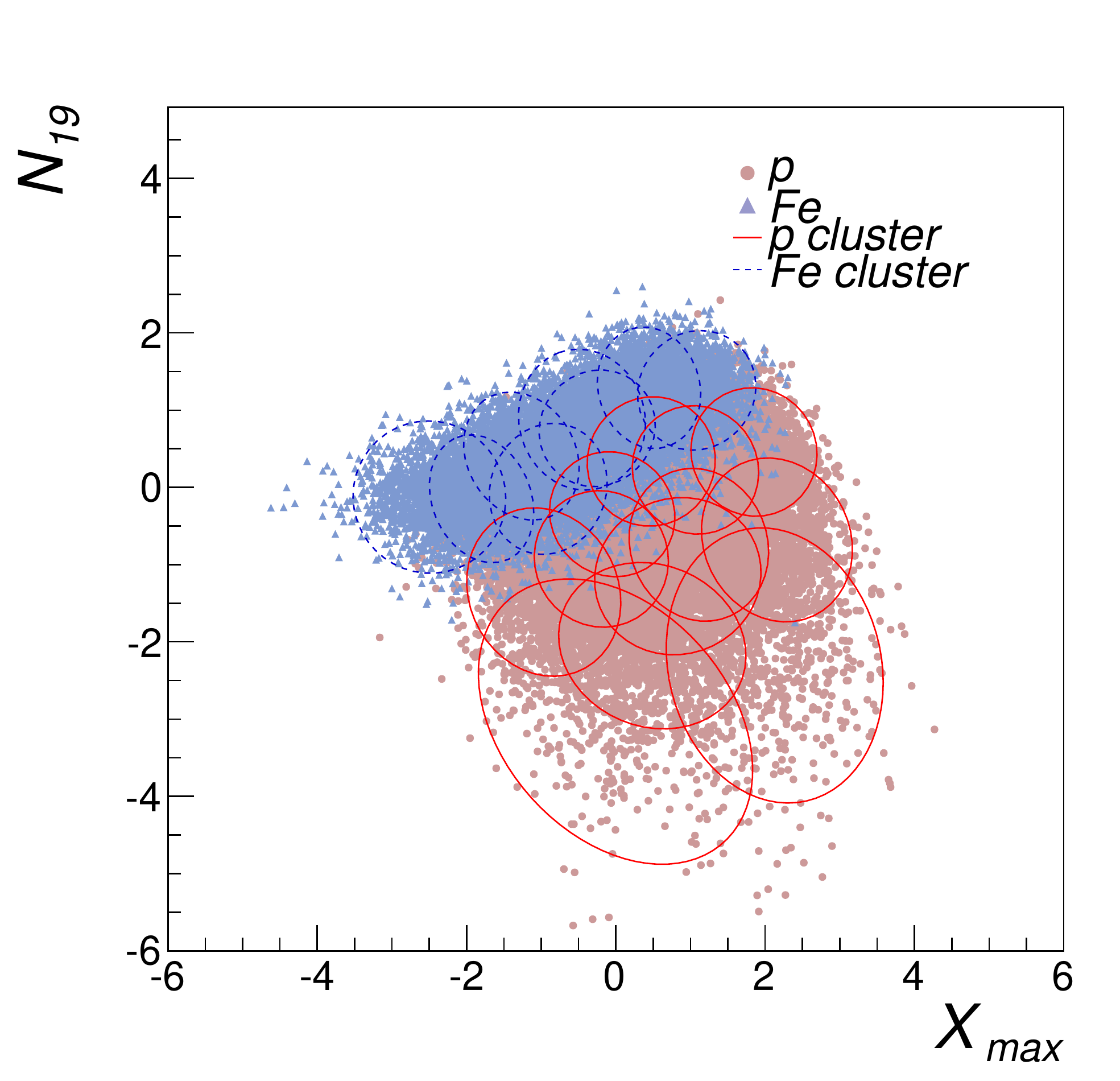}\label{ClusteringOutputFig}}
\subtable[NN output]{\includegraphics[scale=0.3]{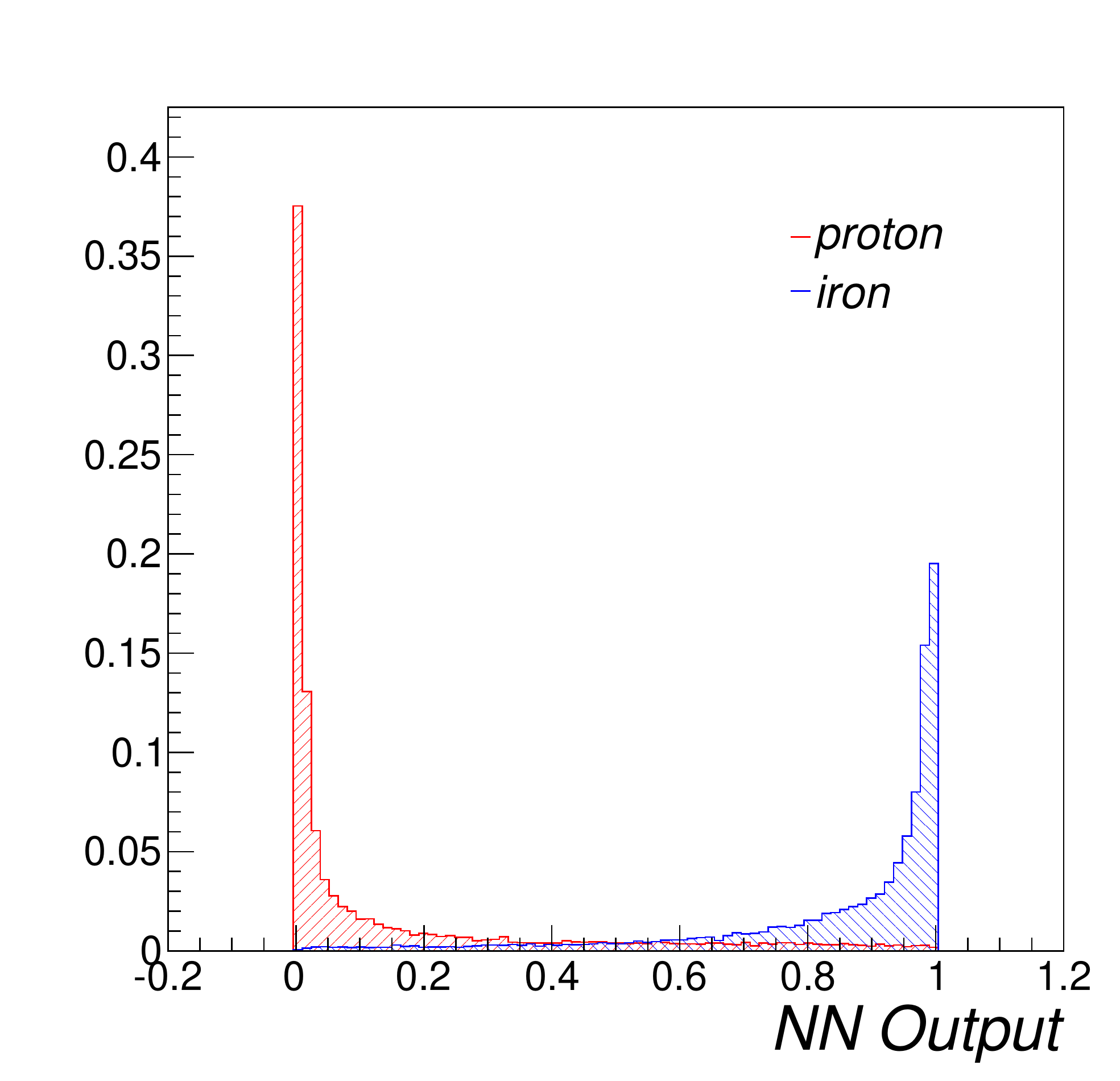}\label{NNOutputFig}}
\caption{\ref{ClusteringOutputFig}: Clustering output projected in the $\Xmax$-$N_{19}$ plane. Proton data are shown with red dots while iron data with blue triangles. 
The ellipses represent the obtained clusters (K=20): proton-labelled clusters (red solid lines), iron-labelled clusters (blue dashed lines). 
Data have been standardized in the plot; \ref{NNOutputFig}: Distribution of neural network outputs in presence of two components, proton (red histogram) and iron 
(blue histogram), for the test data sample. A sample
energy bin corresponding to $\log_{10}$E=[19.0-19.2] is chosen.}%
\label{ClassifierResultFig}
\end{figure*}

\subsection{The Neural Network approach}\label{NNSection}
A feed forward neural network (NN) is structured in parallel layers of neurons, connected to neurons in adjacent
layers through weighted links. The input layer is connected to the input data vector and a predefined number of hidden layers process the signal 
towards the output layer which returns the final response of the network to the presented input data.
Each neuron $i$ linearly transforms the data $x_{j}$ from neurons in the previous layer according to the following expression:
\begin{equation}
z_{i}= \sum_{j=1}^{m}w_{ij}x_{j}+b_{i}
\end{equation}
where $w_{ij}$ are the weights associated to the $j$ link and $b_{i}$ is an additive bias. The neuron output is then obtained by applying a transfer 
function $f(z_{i})$ to the neuron input $z_{i}$.\\During the training phase the network weights and biases are iteratively adjusted to minimize 
the following error function $E_{\textsc{mse}}$:  
\begin{equation}
E_{\textsc{mse}}= \frac{1}{2}\sum_{i=1}^{N}[y_{i}(\textbf{x},\textbf{w})-t_{i}]^{2}
\end{equation}
defined as the sum of squared differences between the desired output $y_{i}$ and the computed network output $t_{i}$.
To improve the network generalization capabilities a regularization term is often added to the above error function to minimize the squared sum of the 
network weights: 
\begin{equation}
E_{\textsc{mse}}^{\textsc{reg}}= \gamma\,E_{\textsc{mse}}+(1-\gamma)\sum_{i=1}^{N_{W}}w_{i}^{2}
\end{equation}
where $N_{W}$ are the number of weights of the network and $\gamma$ is a parameter responsible to guarantee a
compromise between the network error minimization and the
generalization performances of the network.
\\After testing several network architectures and choices 
of transfer functions, we obtained good results using a simple network
design with one hidden layer (3 neurons), and an output layer with one neuron. 
The activation functions are hyperbolic tangent in the hidden layer
and linear in the output layer.\\
The network input data have been normalized to zero mean and 
unit variance and then divided into three subsets, train-validation-test samples. 
The first is used to train the network. The cross validation
sample is used to stop the training at a given epoch to avoid
over fitting the data sample used for learning. The last 
sample is finally used to test the trained network 
identification capabilities.\\Several back propagation training
algorithms have been tested (steepest descent, 
conjugated gradient and quasi-Newton algorithms). 
We achieved better identification performances with quasi-Newton methods using the Broyden-Fletcher-Goldfarb-Shanno (BFGS)
error minimization formula \cite{Bishop,Matlab}.

\begin{figure*}[!th]
\centering%
\subtable[Efficiency]{\includegraphics[scale=0.3]{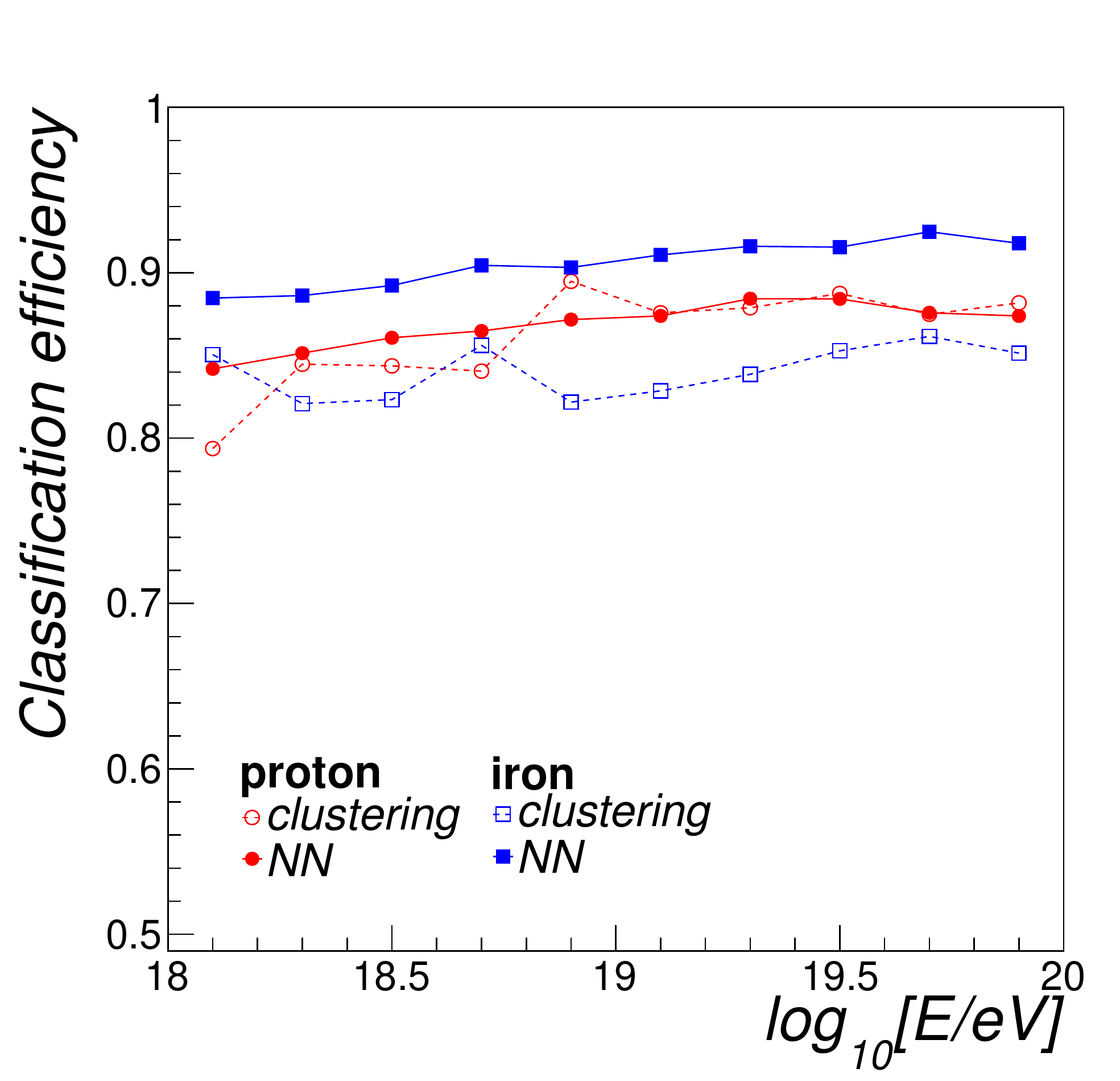}}%
\subtable[Purity]{\includegraphics[scale=0.3]{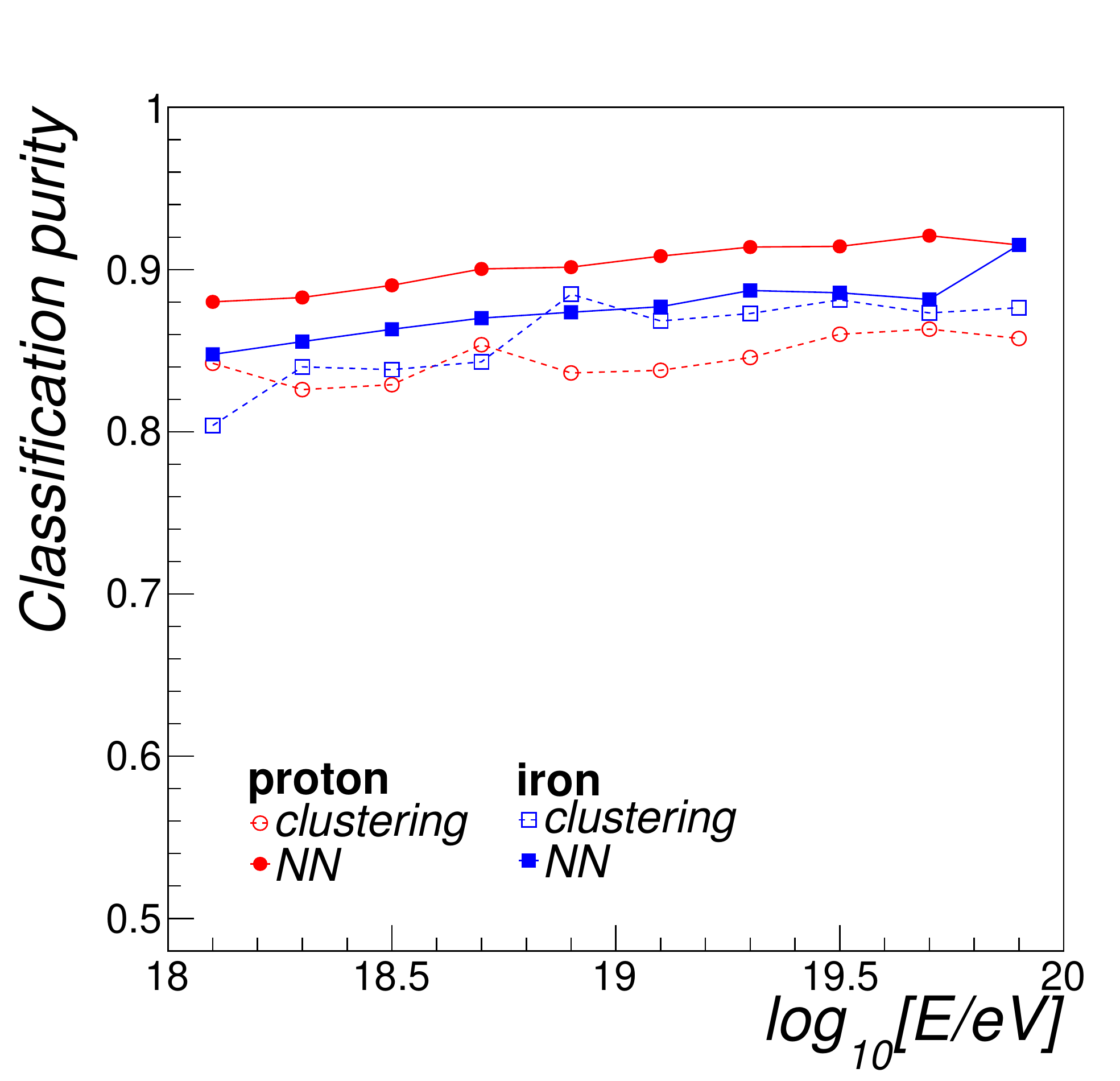}}%
\caption{Classification efficiency (left panel) and purity (right panel) for proton and iron events shown respectively with red dots and blue squares. Results relative to 
the clustering method are shown with empty markers while those obtained with the neural network with filled markers.}
\label{ClassPerformanceVSEnergy}
\end{figure*}

\section{Results}\label{ResultsSection}
\begin{figure*}[!th]
\centering%
\subtable[NN - p+Fe data]{\includegraphics[scale=0.3]{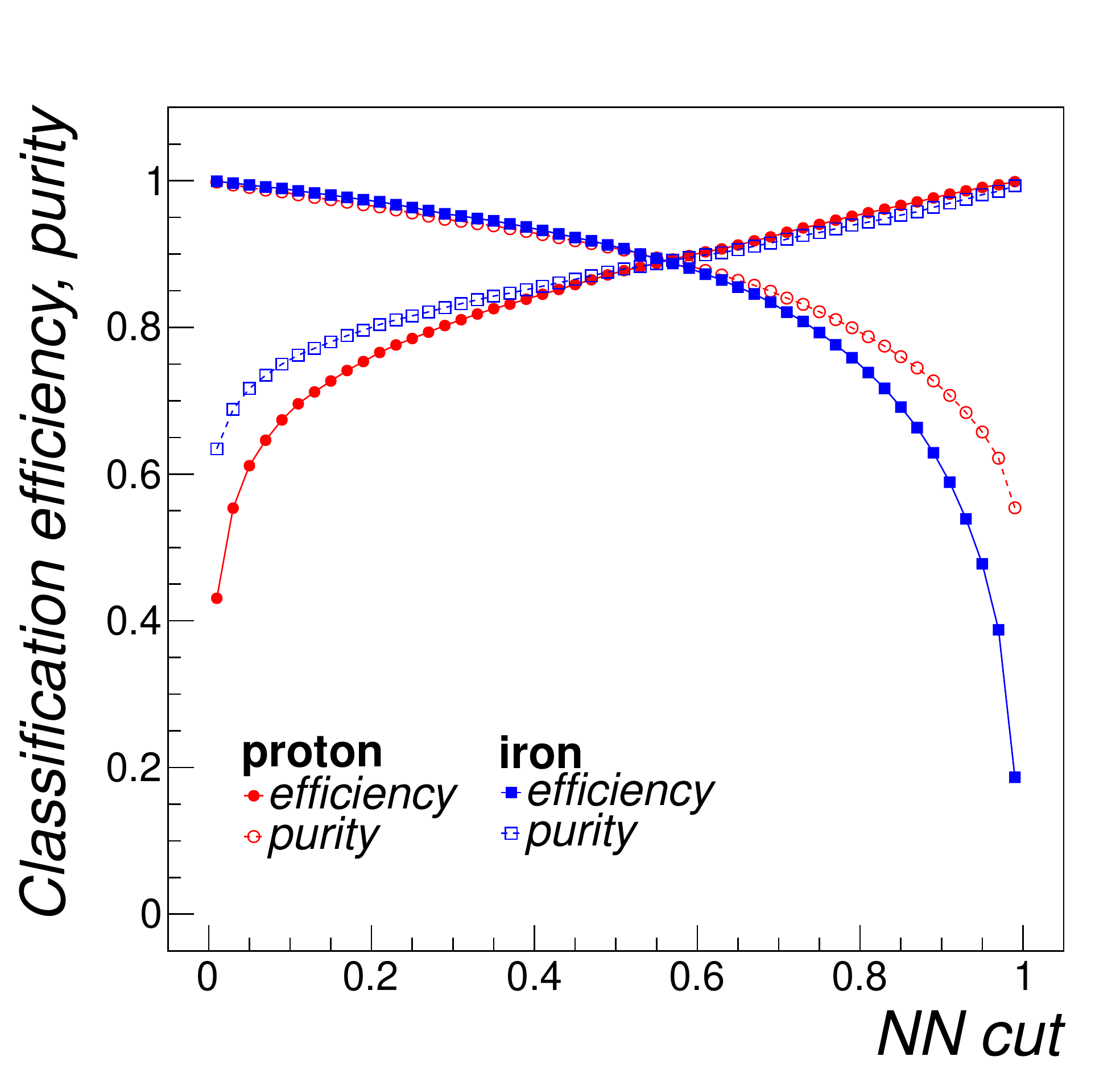}\label{NNPerformanceVSCutFig1}}%
\subtable[NN - p+N+Fe data]{\includegraphics[scale=0.3]{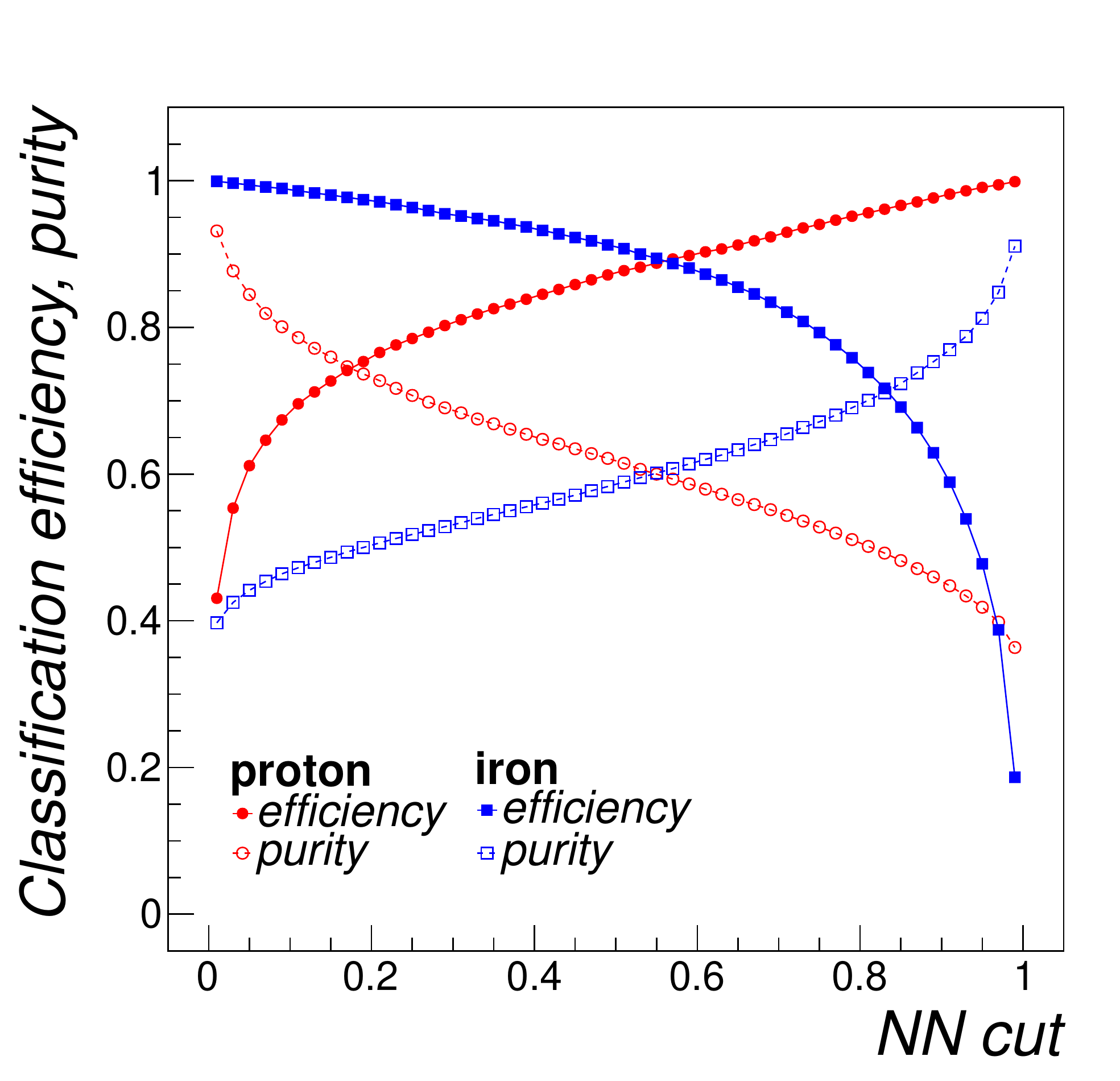}\label{NNPerformanceVSCutFig2}}\\%

\subtable[Clustering - p+Fe data]{\includegraphics[scale=0.3]{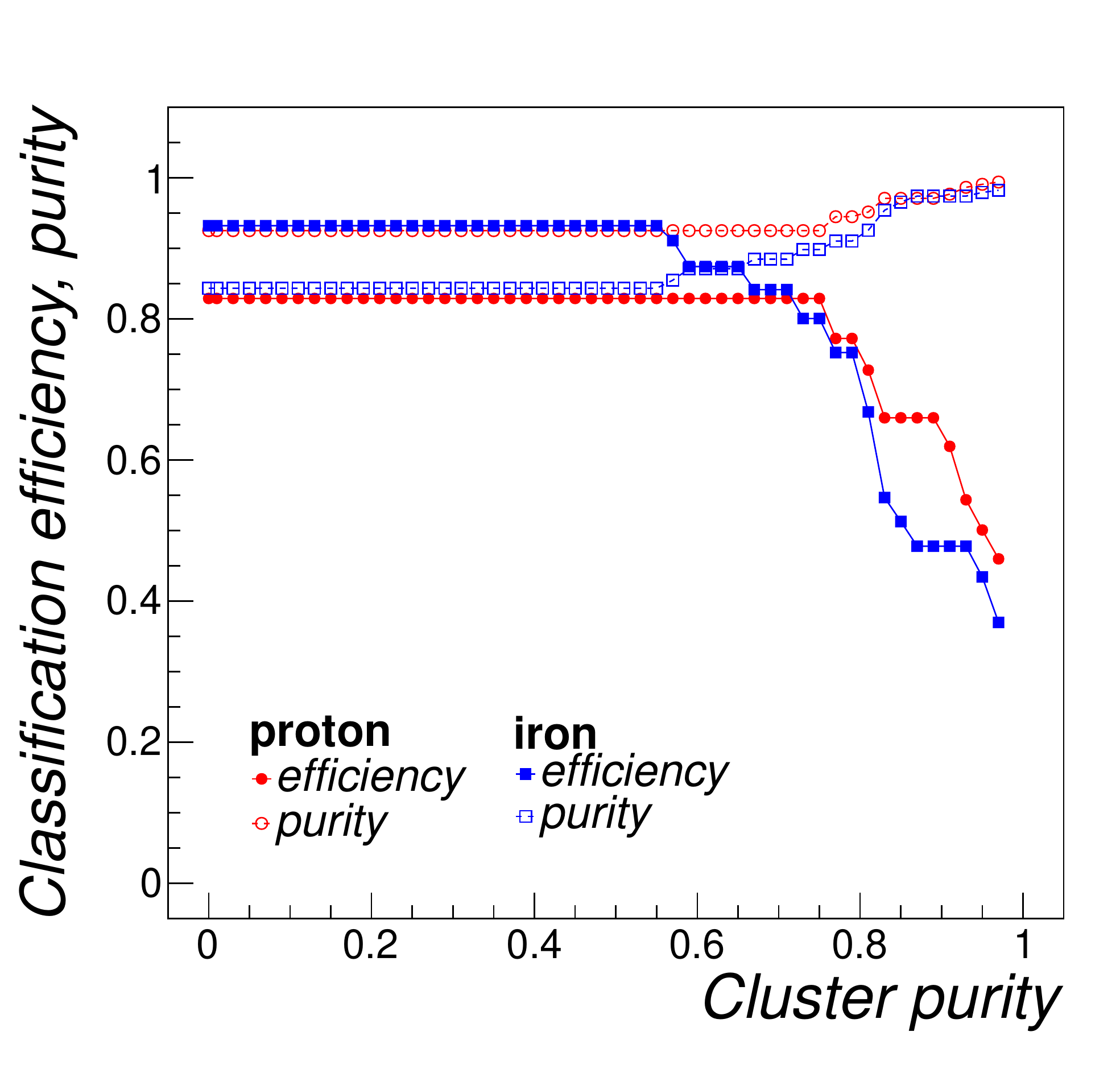}\label{NNPerformanceVSCutFig3}}%
\subtable[Clustering - p+N+Fe data]{\includegraphics[scale=0.3]{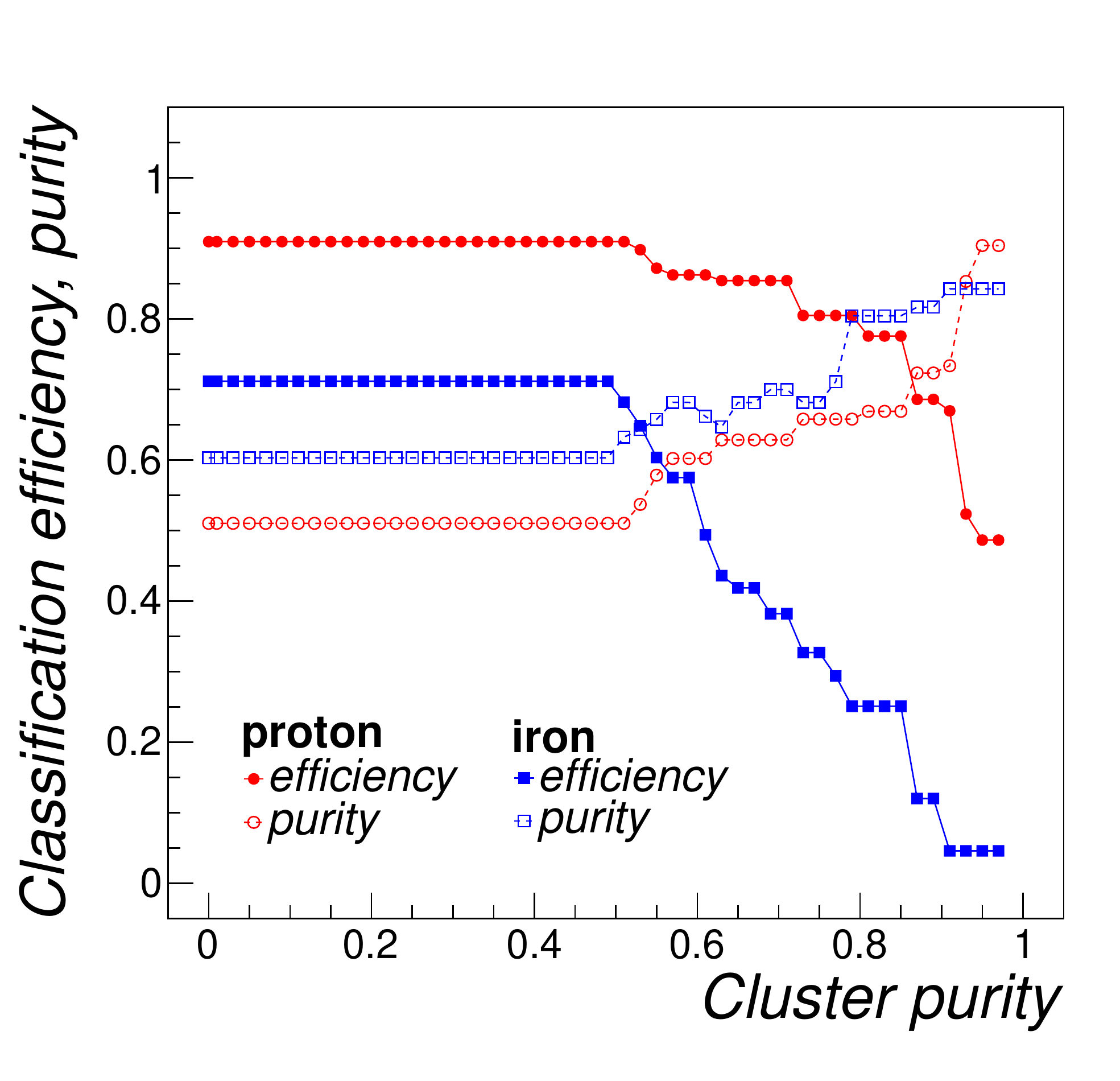}\label{NNPerformanceVSCutFig4}}
\caption{Top panels: Classification efficiency (filled markers) and purity (empty markers) as a function of the cut applied in the neural network output. Proton and 
iron results are respectively shown with red dots and blue squares. Fig. \ref{NNPerformanceVSCutFig1} refers to a 2-components flux made by proton and iron, while in
Fig. \ref{NNPerformanceVSCutFig2} nitrogen is added to the data sample; Bottom panels: Classification efficiency (filled markers) and purity (empty markers) 
as a function of the cluster purity for the clustering classifier method.}
\label{NNPerformanceVSCut}
\end{figure*}

The classification results obtained both with the clustering and the neural network methods are reported in Figs. \ref{ClusteringOutputFig} and \ref{NNOutputFig}
in a sample energy range $\log_{10}$E= [19-19.2].\\In Fig. \ref{ClusteringOutputFig} we report the output of the clustering procedure for the test data sample
projected in the $\Xmax$-$N_{19}$ plane. Proton data are shown with red dots, while iron data with blue triangles. A number of K=20 clusters, 
represented by the ellipses, is considered for this case. A standard representation is adopted to visualize the obtained clusters. First a principal component analysis
is applied to the standardized data to project data in the bidimensional space of the two most important components. The cluster ellipse's axes are then defined by 
the eigenvalues of the covariance matrix 
of the observations belonging to that cluster. Finally the axes are scaled to create a 95\% confidence ellipse.
The achieved classification performances in terms of efficiency and purity are larger than 80\% and 
are reported in Table \ref{ClusteringPerformanceTable}.\\In Fig. \ref{NNOutputFig} we report the distribution of the network outputs in presence of the proton 
(red lines) and iron (blue lines) test data. The performance in terms of classification matrix and purity relative to such case are reported in 
Table \ref{NNPerformanceTable}. Assuming a cut in the network output
equal to 0.5, the classification efficiencies and purity are found to
be around 90\%.
\begin{table}[!h]
\centering
%\footnotesize%
\begin{tabular}{|c|ccc}
 \hline%
 & \multicolumn{2}{c}{\textbf{Classification}} & \multicolumn{1}{|c|}{\textbf{Purity}}\\
 & \multicolumn{1}{c}{\textit{p}} & \multicolumn{1}{c}{\textit{Fe}} & \multicolumn{1}{|c|}{}\\
 \cline{1-4}%
 \textit{p}  & 0.84 & 0.10 & \multicolumn{1}{|c|}{0.91}\\%
 \textit{Fe} & 0.16 & 0.90 & \multicolumn{1}{|c|}{0.88}\\% 
 \hline
\end{tabular}
\caption{Classification matrix and purity for the clustering method for the sample test data in the energy bin $\log_{10}E$= [19.0,19.2].}%
\label{ClusteringPerformanceTable}
\end{table}
\\In Fig. \ref{ClassPerformanceVSEnergy} we report the classification
efficiency for proton and iron (left panel) and the purity (right panel)
as a function of the primary energy. Results obtained with the neural
network (clustering) classifier are shown with filled (empty) dots. Both methods are able to 
accurately discriminate the two presented classes over the entire data 
set. The neural network method performs slightly better than 
the clustering approach. As expected, the classification performances
are found to slightly increase 
with energy due to the improved reconstruction resolution for 
both $\Xmax$ and $N_{19}$ parameters.
\begin{table}[!h]
\centering
%\footnotesize%
\begin{tabular}{|c|ccc}
\hline%
& \multicolumn{2}{c}{\textbf{Classification}} & \multicolumn{1}{|c|}{\textbf{Purity}}\\
& \multicolumn{1}{c}{\textit{p}} & \multicolumn{1}{c}{\textit{Fe}} & \multicolumn{1}{|c|}{}\\
\cline{1-4}%
\textit{p}  & 0.87 & 0.09 & \multicolumn{1}{|c|}{0.83}\\%
\textit{Fe} & 0.13 & 0.91 & \multicolumn{1}{|c|}{0.86}\\% 
\hline
\end{tabular}
\caption{Classification matrix and purity for the neural network method for the sample test data in the energy bin $\log_{10}E$= [19.0,19.2].}%
\label{NNPerformanceTable}
\end{table}
\\To have the smallest
contamination for a given class and match a given analysis requirement one can tighten the applied
selection cuts, increasing at the same time the event rejection rate,
as shown in Figs. \ref{NNPerformanceVSCutFig1} and \ref{NNPerformanceVSCutFig3}. 

In Fig. \ref{NNPerformanceVSCutFig1} we report the classification
efficiency (filled dots) and purity (empty dots) as a function of the
network output for proton and iron data. The best compromise between the highest 
efficiency and purity is found around 0.55-0.60. To decrease the contamination from other species below few percent a cut of $\sim$0.2 for proton and $\sim$0.8 for iron
must be assumed.\\Similarly, for the clustering method a smaller contamination can be achieved by rejecting those clusters having a classification purity below a 
given requirement during the training phase. We therefore report in Fig. \ref{NNPerformanceVSCutFig3} the results as a function of the cluster purity obtained
for the training data set. The steps observed in the plots are due to the fact that for a given purity threshold an entire set of events belonging to a cluster 
below threshold are rejected. The effect is more pronounced when a small number of clusters is assumed in the analysis.
\\The presented results so far assumed that no other species other than proton and iron is present in the flux. If an intermediate component, 
like nitrogen, is included in the data and both classifier are trained to recognize three species, the number of classifier parameters needed to achieve optimal results 
must be increased. The obtained performances deteriorate significantly in a 3-component situation. 

For the extreme components the identification efficiency and purity
drops to a level of 75\%, while 
for the intermediate component it is of order 60\%. It is clear that with
current parameter sensitivity to the mass and reconstruction
resolution it is not possible to 
efficiently discriminate more than two masses. We therefore quantify
in Figs. \ref{NNPerformanceVSCutFig2}, \ref{NNPerformanceVSCutFig4}
the increased contamination 
in the proton and iron classified samples due to the presence of nitrogen.
For the neural network an optimal compromise is found assuming a cut
equal to 0.2 and 0.8. The 
corresponding efficiency and purity are reduced at the 80\% level. However
one can obtain smaller contamination by further tightening the applied
cuts at the cost of reducing the 
detection efficiency. For the clustering method the presence of 
an intermediate component rises the contamination in the proton and
iron sample to values around 40-50\%. 
To recover the contamination observed with proton and iron 
components only a large number of clusters must be rejected. 

\section{Summary}\label{SummarySection}
In the present work we propose two alternative strategies to the
problem of the high energy cosmic ray 
mass identification on a event-by-event basis. One is based on
the neural network technique while the other on clustering
algorithms. The performances of 
both classifiers have been studied with simulated showers generated
with the \textsc{Conex} tool. The mass discrimination is based on two
parameters, the depth of 
shower maximum $\Xmax$ and an estimator of the muon number 
$N_{19}$ reconstructed in very inclined showers. Realistic
reconstruction resolutions have been assumed for both parameters. 
The analysis is in particular optimized for events recorded by hybrid
detectors with zenith angles above 
60$^{\circ}$. Nevertheless it can be extended in a 
straightforward way to nearly vertical events, provided that an
alternative muon number parameter 
estimator is introduced.\\We have found very good identification 
performances for both methods in the case of a 2-component flux of 
proton and iron nuclei. The expected misclassification are found of the order of $\sim$10-15\%, 
decreasing with energy, with slightly better performances exhibited by
the neural network classifier.
\\A significant loss of efficiency is observed when training 
the classifiers to recognize a flux with an intermediate mass component
added. In this situation the 
expected contamination affecting the proton and iron identification 
has been reported. No significant performance degradation has been
observed with an alternative data 
sample generated using a different hadronic interaction model, 
such as \textsc{Epos}.
\\The present method, as it is, can be applied also to other typical
discrimination 
problems in cosmic ray physics, such as hadron-gamma 
or hadron-neutrino event discrimination. In such cases the
classification performances are 
expected to increase given the better separation between the two classes. 
Other feasible applications include the selection of a given mass
group from the background, as it could 
be done for the selection of pure samples of 
protons for correlation 
analysis with astrophysical objects or to derive estimates of the
proton-Air cross section.
\\In this work we made use of supervised methods. 
An implicit assumption is made, namely that the measured data must be well 
described or at least bracketed by the hadronic model predictions adopted in the classifiers. 
Surprisingly, this is the case for most of the shower observables so far measured at such extreme 
energies, i.e. see \cite{UngerKampertReview}. 
Furthermore, the differences among different models are going to be
significantly reduced with incoming model 
releases accounting for recent LHC measurements, 
for instance see \cite{PierogECRC2012}. Recently a significant
discrepancy 
between data and Monte Carlo simulations has been reported in the muon number estimator both for 
vertical \cite{AllenICRC2011} and for very inclined showers
\cite{GonzaloICRC2011}. 
For this reason we are currently developing a semi-supervised classifier, 
partially using the strong constraints offered by the available
models. Results and performance comparison with supervised methods will be reported in a 
forthcoming paper.
 
%Provided that the measured data are bracketed by the Monte Carlo data adopted in the classifiers, the developed analysis can be applied to
%to select pure data samples of protons for correlation analysis with astrophysical objects or to derive estimates of the 
%proton-Air cross section.  

\section{Acknowledgements}
S. Riggi acknowledges the astroparticle group of the Universidad de
Santiago de Compostela for support and ideas exchanging.
We thank Xunta de Galicia (INCITE09 206 336 PR) and 
Conseller\'\i a de Educaci\'on (Grupos de Referencia Competitivos -- 
Consolider Xunta de Galicia 2006/51); Ministerio de  Educaci\'on,
Cultura y Deporte, Spain (FPA 2010-18410, FPA 2012-39489 and 
Consolider CPAN - Ingenio 2010) and Feder Funds.
We are greatful to CESGA (Centro de SuperComputaci\'on de Galicia) for computing resources.


\begin{thebibliography}{99}
\bibitem{UngerKampertReview} K.H. Kampert, M. Unger, Astroparticle Physics 35 (2012), 660.%
\bibitem{DUrsoICRC09} D. D'Urso for the Pierre Auger Collaboration, Proc. of 31st International Cosmic Ray Conference (2009).
\bibitem{KASCADEAnalysis} T. Antoni at al, Astroparticle Physycs 24 (2005) 1; 
W.D. Apel et al, Astroparticle Physics 31 (2009) 86.
\bibitem{Tiba} A.K.O. Tiba, G.A.Medina-Tanco, S.J.Sciutto, \emph{astro-ph0502255}.%
\bibitem{Ambrosio} M.Ambrosio et al, Astroparticle Physics 24 (2005) 355.%
\bibitem{NeutrinoAugerPaper} P. Abreu et al, Physical Review D 84 (2011) 122005.%
\bibitem{PAODesignReport} J. Abraham et al, Nucl. Instr. Meth. A 523 (2004) 50.
\bibitem{TADesignReport} T. Nonaka for the Telescope Array Collaboration, Proc. of 31st International Cosmic Ray Conference (2009); 
A. Taketa for the Telescope Array Collaboration, Proc. of 31st International Cosmic Ray Conference (2009); 
Y. Tameda et al, Nucl. Instr. Meth. A 609 (2009) 227.
\bibitem{GonzaloICRC2011} G. Rodriguez for the Pierre Auger Collaboration, Proc. of 32nd International Cosmic Ray Conference (2011).%
\bibitem{UngerProfileReco} M. Unger, B.R. Dawson, R. Engel, F. Schussler and R. Ulrich, Nucl. Instr. Meth. A 588 (2008) 433.%
\bibitem{Conex1} T. Pierog et al, Nucl. Phys. B (Proc. Suppl.) 151 (2006) 159.%
\bibitem{Conex2} N.N. Kalmykov et al, Astroparticle Physics 26 (2007) 420.%
\bibitem{QGSJET01Paper} N.N. Kalmykov et al, Nucl. Phys. B Proc. Suppl. 52 (1997) 17.%
\bibitem{CORSIKAPaper} D. Heck, J. Knapp, J.N. Capdevielle, G. Schatz, T. Thouw, Forschungszentrum Karlsruhe Report FZKA 6019 (1998).%
\bibitem{AIRESPaper} S.J. Sciutto, Proc. of 27th International Cosmic Ray Conference (2001).%
%\bibitem{FacalICRC11} P. Facal for the Pierre Auger Collaboration, Proc. of 32nd International Cosmic Ray Conference (2011).%
\bibitem{XmaxPRL} J. Abraham et al, Phys. Rev. Letters 104 (2010) 091101.
\bibitem{YounkRisse} P. Younk and M. Risse, Astroparticle Physics 35 (2012) 807.%
\bibitem{KMeansPaper} J.A. Hartigan, M.A. Wong, Journal of the Royal Statistical Society C 28 (1979) 100.% 
\bibitem{Bishop} C.M. Bishop, Neural Networks for Pattern Recognition, Oxford University Press (1995).%
\bibitem{Matlab} H. Demuth, M. Beale, Neural Network Toolbox - MATLAB - User's Guide (2006), version 5.%
\bibitem{PierogECRC2012} T. Pierog, Proc. of 23rd European Cosmic Ray Symposium (2012).%
\bibitem{AllenICRC2011} J. Allen for the Pierre Auger Collaboration, Proc. of 32nd International Cosmic Ray Conference (2011).%

\end{thebibliography}
\end{document}